\documentclass{article}

\usepackage{amsthm,amsmath,amsfonts}
\RequirePackage[colorlinks,citecolor=blue,urlcolor=blue]{hyperref}
\usepackage[round]{natbib}
\usepackage{bm}
\usepackage{graphicx}
\usepackage{booktabs}
\usepackage{geometry}
\newcommand{\tp}{^{\top}}
\newcommand{\rd}{\mathrm{d}}
\newcommand{\rds}{\,\rd}
\DeclareMathOperator{\expect}{\mathrm{E}}
\DeclareMathOperator{\var}{\mathrm{var}}
\def\eqd{\,{\buildrel  \mathcal{D} \over =}\,}
\def\iid{\,{\buildrel  \mathrm{iid} \over \sim}\,}
\DeclareMathOperator*{\argmax}{arg\,max}

\title{Estimation of the generalized Laplace distribution and its projection onto the circle}
\author{Marco Geraci\\Sapienza University of Rome}
\date{}

\begin{document}

\maketitle

\begin{abstract}
The generalized Laplace (GL) distribution, which falls in the larger family of generalized hyperbolic distributions, provides a versatile model to deal with a variety of applications thanks to its shape parameters. The elliptically symmetric GL admits a polar representation that can be used to yield a circular distribution, which we call \emph{projected} GL (PGL) distribution. The latter does not appear to have been considered yet in practical applications. In this article, we explore an easy-to-implement maximum likelihood estimation strategy based on Gaussian quadrature for the scale-mixture representation of the GL and its projection onto the circle. A simulation study is carried out to benchmark the fitting routine against expectation-maximization and direct maximum likelihood to assess its feasibility, while the PGL model is contrasted with the von Mises and projected normal distributions to assess its prospective utility. The results showed that quadrature-based estimation is more reliable consistently across selected scenarios and sample sizes than alternative estimation methods, while the PGL complements other distributions in terms of flexibility. \\
\vspace{0.5cm}
\end{abstract}

\section{Introduction}\label{sec:1}
Many are familiar with the classical Laplace distribution. Its probability density function (PDF) is given by
\begin{equation}\label{eq:1}
f_{L}(y) = \frac{1}{\sqrt{2}\sigma} \exp \left\{-\frac{\sqrt{2}}{\sigma}|y-\theta|\right\},
\end{equation}
with $y \in \mathbb{R}$, location $\theta \in \mathbb{R}$ and scale $\sigma > 0$. This distribution, which we denote $\mathcal{L}(\theta,\sigma)$, is symmetric about $\theta$ and is often considered the `robust' counterpart of the Gaussian distribution. This is because the maximum likelihood (ML) estimate of $\theta$ is the sample median \citep{keynes1911}. Its mean and variance are given by $\expect(Y) = \theta$ and $\var(Y) = \sigma^2$, respectively. Moreover, if $Y \sim \mathcal{L}(\theta,\sigma)$, we have the scale-mixture representation $Y \eqd \theta + \sqrt{V}Z$, where $V\sim \mathcal{E}(1)$ is standard exponential and $Z \sim \mathcal{N}(0,\sigma^2)$ is normal with mean $0$ and variance $\sigma^2$. That is, the Laplace arises from a heterogeneous population of normals which makes it an apposite model in all situations when heterogeneity in the population is suspected and the observations show large errors \citep{geraci2018}.

The asymmetric Laplace (AL) distribution is an extension of \eqref{eq:1} to account for skewness. Its PDF is given by \citep{kotz2001}
\begin{equation}\label{eq:2}
f_{AL}(y)=\frac{1}{\sqrt{2 + \mu^2/\sigma^2}\sigma}
    \begin{cases}
        \exp \left(-\frac{\sqrt{2}}{\sigma}\frac{\sqrt{2\sigma^2 + \mu^2}-\mu}{\sqrt{2}\sigma}|y-\theta|\right), & \mbox{if $y \geq \theta$}\\
        \exp \left(-\frac{\sqrt{2}}{\sigma}\frac{\sqrt{2\sigma^2 + \mu^2}+\mu}{\sqrt{2}\sigma}|y-\theta|\right), & \mbox{if $y < \theta$}
    \end{cases}
\end{equation}
where $\mu \in \mathbb{R}$ is the asymmetry parameter. The density in~\eqref{eq:2} reduces to~\eqref{eq:1} for $\mu = 0$. The AL, denoted here $\mathcal{AL}(\theta,\sigma,\mu)$, has been further generalized to include an additional shape parameter, say $\alpha > 0$, which controls the mode and the tails of the distribution, and has PDF
\begin{equation}\label{eq:3}
f_{GL}(y)=\frac{2 \exp\left\{\mu|y-\theta|\sigma^{-2}\right\}}{\sqrt{2\pi} \Gamma(\alpha)\sigma}\left(\frac{|y-\theta|\sigma^{-1}}{\sqrt{2 + \mu^2/\sigma^2}}\right)^{\alpha-1/2} K_{\alpha-1/2}\left(\sqrt{2 + \mu^2/\sigma^2}|y-\theta|\sigma^{-1}\right), \qquad y \neq \theta,
\end{equation}
where $K_{u}$ is the modified Bessel function of the third kind with index $u$. The generalized Laplace (GL) in~\eqref{eq:3} (also known as variance-gamma distribution), denoted hereafter $\mathcal{GL}(\theta, \sigma, \mu, \alpha)$, coincides with the AL in~\eqref{eq:2} when $\alpha = 1$. (To see this, note that $K_{1/2}(x) = \sqrt{\pi/(2x)}e^{-x}$.) The mean and variance of the univariate GL are given by, respectively, $\expect(Y) = \theta + \alpha\mu$ and $\var(Y) = \alpha(\sigma^2 + \mu^2)$ \citep{kozu2013}. Interestingly, a GL random variable converges in distribution to a Gaussian variable when $\alpha \rightarrow \infty$ (provided that $\alpha\mu_{\alpha}$ and $\alpha\sigma_{\alpha}^2$ converge to finite values as $\alpha$ diverges to infinity) \citep[see][p.183, for further details]{kotz2001}. The GL distribution is a limiting case of the generalized hyperbolic (GH) distribution \citep{barndorff1977}.

We now define the multivariate extension of \eqref{eq:3}. An $d$-dimensional random variable $\bm{Y} = (Y_{1}, Y_{2}, \ldots, Y_{d})\tp$ is said to follow a multivariate GL distribution if it has PDF
\begin{equation}\label{eq:4}
f_{GL}(\bm y) = \dfrac{2\exp\{\bm{\mu}\tp \bm\Sigma^{-1}(\bm{y}-\bm\theta)\}}{(2\pi)^{d/2}\Gamma(\alpha)|\bm\Sigma|^{1/2}}\left(\frac{Q(\bm y, \bm\theta, \bm\Sigma)}{P(\bm\Sigma,\bm\mu)}\right)^{\alpha - d/2}K_{\alpha - d/2}\left(Q(\bm y, \bm\theta, \bm\Sigma)P(\bm\Sigma,\bm\mu)\right), \qquad \bm{y} \neq \bm{\theta},
\end{equation}
where $\bm{y} \in \mathbb{R}^{d}$, $\bm\theta \equiv (\theta_{1}, \ldots, \theta_{d})\tp \in \mathbb{R}^{d}$, $\bm\Sigma$ is a positive-definite $d \times d$ matrix and has generic element $\sigma_{ij}$, $1\leq i \leq j \leq d$, $\bm{\mu} \equiv (\mu_{1}, \ldots, \mu_{d})\tp \in \mathbb{R}^{d}$, $\alpha > 0$, $Q(\bm y, \bm\theta, \bm\Sigma) = \sqrt{(\bm y-\bm\theta)\tp \bm\Sigma^{-1}(\bm y-\bm\theta)}$, and $P(\bm\Sigma,\bm\mu) = \sqrt{2 + \bm{\mu}\tp \bm{\Sigma}^{-1}\bm{\mu}}$. This distribution is denoted by $\mathcal{GL}_{d}(\bm\theta,\bm\Sigma,\bm\mu,\alpha)$. The mean and variance-covariance of the multivariate GL are given by, respectively, $\expect(\bm Y) = \bm\theta + \alpha\bm\mu$ and $\var(\bm Y) = \alpha(\bm\Sigma + \bm{\mu}\bm{\mu}\tp)$ \citep{kozu2013}. Note that these expressions reduce to those given for the univariate GL in~\eqref{eq:3} when $d = 1$ and for the symmetric Laplace in~\eqref{eq:1} when $d = 1$, $\mu = 0$ and $\alpha = 1$. Similarly to~\eqref{eq:3}, the multivariate GL in~\eqref{eq:4} converges to a multivariate normal as $\alpha$ becomes large (again, as long as the mean and variance converge to finite values). Also, $\mathcal{GL}_{d}(\bm\theta, \bm\Sigma, \bm\mu, 1)$ represents a multivariate generalization of the AL in $\eqref{eq:2}$.

Like other elliptical distributions, the symmetric GL ($\bm{\mu} = \bm{0}$) has a polar representation \citep{kozu2013}, namely
\begin{equation}\label{eq:5}
  \bm Y \eqd \bm{\theta} + R \bm H \bm U,
\end{equation}
where $\bm Y \sim \mathcal{GL}_{d}(\bm{\theta},\bm\Sigma,\bm{0}_{d},\alpha)$, $\bm{H}$ is a $d \times d$ matrix such that $\bm{H} \bm{H}\tp = \bm{\Sigma}$, $\bm{U}$ is a random vector uniformly distributed on the unit sphere $\mathbb{S}_{d}$ of $\mathbb{R}^{d}$, and $R$ is a positive random variable, independent of $\bm{U}$, with PDF
\begin{equation}\label{eq:6}
f_{R}(r)=\dfrac{2 r^{d / 2+\alpha-1} K_{-\alpha + d/2}(\sqrt{2} r)}{(\sqrt{2})^{d / 2+\alpha-2} \Gamma(\alpha) \Gamma(d / 2)}.
\end{equation}

\begin{figure}[h]
    \centering
    \includegraphics[scale=0.6]{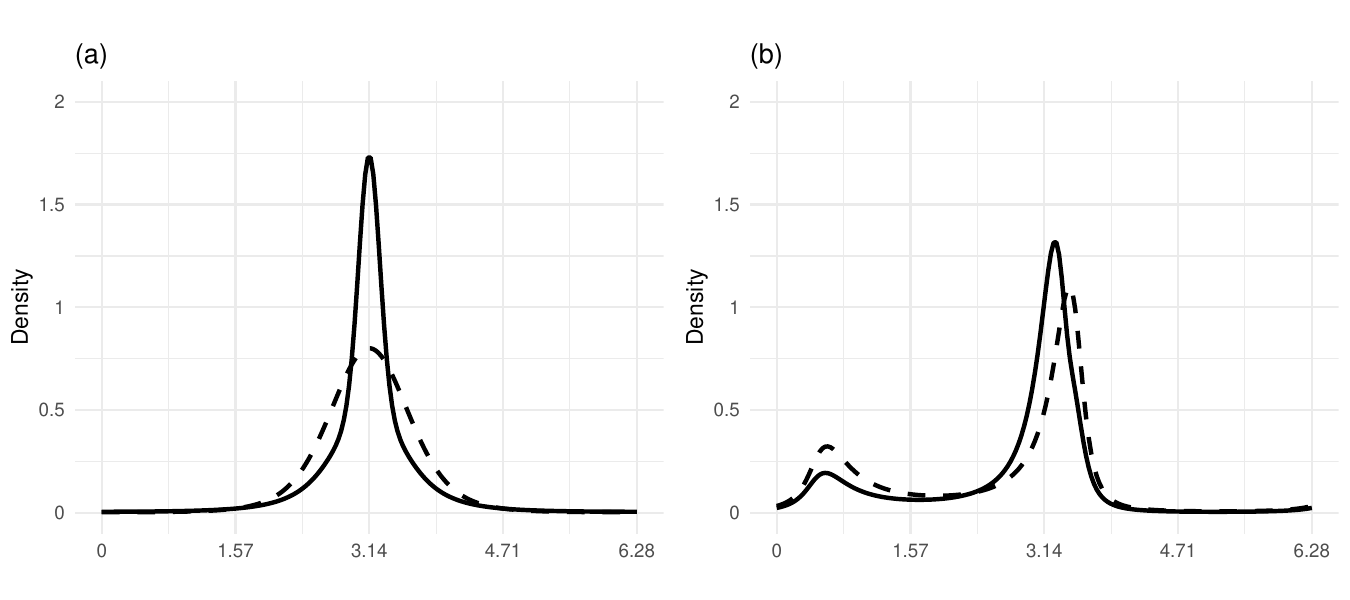}
    \caption{Examples of (a) unimodal and (b) bimodal projected generalized Laplace densities with $\alpha = 0.5$ (solid lines) and $\alpha = 10$ (dashed lines). The latter is approximately a projected normal.}
    \label{fig:1}
\end{figure}

A practical application of \eqref{eq:5} is in circular statistics. Let $\bm{S}= R\bm{W}$, where $\bm{W} = (\cos\Omega,\sin\Omega)\tp$. Assume $\bm{S} \sim \mathcal{GL}_{2}(\bm{\theta},\bm\Sigma,\bm{0}_{2},\alpha)$. We define the projected generalized Laplace (PGL) density for the random variable $\Omega$
\begin{equation}\label{eq:7}
f_{PGL}(\omega) = \int_{0}^{\infty} |\bm{J}| f_{GL}(r\bm{w}) \rds r = \int_{0}^{\infty} \dfrac{r|\bm\Sigma|^{-1/2}}{\pi\Gamma(\alpha)2^{(\alpha-1)/2}}\left(Q(r\bm{w}, \bm{\theta}, \bm{\Sigma})\right)^{\alpha - 1}K_{\alpha - 1}\left(\sqrt{2}Q(r\bm{w}, \bm{\theta}, \bm{\Sigma})\right) \rds r,
\end{equation}
where $\omega \in (-\pi,\pi]$, $|\bm{J}| = r$ is the determinant of the Jacobian of the transformation $(s_{1},s_{2}) \rightarrow r(\cos\omega,\sin\omega)$ and $\bm{w} = (w_{1},w_{2})\tp$. This distribution, denoted by $\mathcal{PGL}(\bm{\theta},\bm\Sigma,\alpha)$, is akin to the projected normal (PN) \citep{wang2013}. Indeed, the latter is a special case of the PGL for $\alpha \rightarrow \infty$. Both the PGL and the PN can be symmetric, asymmetric, unimodal or bimodal. However, compared to the PN, the PGL has the additional shape parameter $\alpha$ that controls the spikedness of the main mode (Figure~\ref{fig:1}). The PGL (like the PN) is therefore a flexible alternative to the much celebrated von Mises (VM) distribution, a unimodal, symmetric model that has a long history in circular statistics. Finally, our PGL is a generalization of the projected Laplace in \cite{siew2008} for general location and scale parameters.

Direct ML estimation of the parameters in~\eqref{eq:4} and~\eqref{eq:7} poses computational challenges due to the complex features of these densities such as the presence of the Bessel function, lack of differentiability and irregular behavior near $\bm{\theta}$. (This is true for GH distributions in general \citep{eberlein1995,vanwyk2024}.) In addition, the integral in~\eqref{eq:7} does not appear to have a closed-form solution. On the other hand, working with the scale-mixture representation of the GL opens up alternative strategies that may be computationally attractive. Among these, one strategy that is popular with GH-family models is the expectation-maximization (EM) algorithm \citep{karlis2002,aas2006,hellmich2011}. Gaussian quadrature (GQ) is another strategy that has comparatively received less attention despite being relatively easy to implement and to adapt to non-standard situations. See for example \cite{geraci2020} for GL mixed-effects models using GQ and \cite{choi2021} for numerical approximations of the GH distribution and related moments.

\section{Estimation}\label{sec:2}
For $\bm{Y} \sim \mathcal{GL}_{d}(\bm\theta,\bm\Sigma,\bm\mu,\alpha)$, we have the representation \citep{kozu2013}
\begin{equation}\label{eq:8}
  Y \eqd \bm{\theta} + V\bm{\mu} + \sqrt{V} \bm{Z},
\end{equation}
where $V\sim \mathcal{G}(\alpha,1)$ is a standard gamma with shape $\alpha$ and $Z\sim\mathcal{N}_{d}(\bm{0}_{d},\bm{\Sigma})$. For $d=1$, $\mu = 0$ and $\alpha = 1$ (hence $\mathcal{G}(1,1)$ is standard exponential), we obtain the scale-mixture representation of the Laplace. We use the scale-mixture representation in~\eqref{eq:8} to obtain the likelihood for a sample $\bm{Y} = (\bm{Y}_{1},\ldots,\bm{Y}_{n}) \iid \mathcal{GL}_{d}(\bm\theta,\bm\Sigma,\bm\mu,\alpha)$. Note that the mixing variable $V$ is latent, hence it needs to be integrated out. Let $\bm{\delta}$ be the unrestricted vector of parameters in $\bm{\Sigma}$ using the matrix-logarithm parametrization described in \cite{pinheiro1996}, and $\alpha = \exp(\zeta)$. Then, the parameter to be estimated is $\bm{\eta} = (\bm{\theta}\tp, \bm{\delta}\tp, \bm{\mu}\tp, \zeta)\tp \in \mathbb{R}^{k}$, where $k = d(d + 5)/2+1$. We have the loglikelihood
\begin{equation}\label{eq:9}
\ell(\bm{\eta}; \bm{y}) = \sum_{i=1}^{n} \log \int_{0}^{\infty} f_{N}(\bm{y}_{i}|v)g(v) \rds v,
\end{equation}
where $f_{N}(\bm{y}|v)$ is the density of $\bm{Y}|V \sim \mathcal{N}_{d}(\bm{\theta} + v\bm{\mu},v\bm{\Sigma})$ and $g(v)$ the density of $V\sim \mathcal{G}(\alpha,1)$. By applying a GQ to~\eqref{eq:9}, we aim at solving the following (approximated) ML estimation problem
\begin{equation}\label{eq:10}
\hat{\bm{\eta}} = \argmax_{\bm{\eta}} \sum_{i=1}^{n} \log \left[\sum_{h=1}^{H}  \left\{f_{N}(\bm{y}_{i}|v_{h}(\alpha))\right\}w_{h}(\alpha)\right],
\end{equation}
where $v_{h}(\alpha)$ and $w_{h}(\alpha)$ are, respectively, the node and weight of a gamma quadrature rule, $h = 1,\ldots,H$, with shape parameter $\alpha$. These can be conveniently obtained with \texttt{statmod::gauss.quad.prob(..., dist = "gamma", alpha)} in \texttt{R}. The optimization problem in~\eqref{eq:10} can be carried out with general-purpose optimizers. In our \texttt{R} implementation, we experienced convergence failures with the derivative-free (Nelder-Mead) and the quasi-Newton (BFGS) algorithms in \texttt{optim}. In contrast, the quasi-Newton algorithm in \texttt{nlminb} was more reliable, perhaps due to its automatic scaling of the parameters. Note that any of the densities in~\eqref{eq:1}-\eqref{eq:4} can be fitted with the procedure described above. Of course, more efficient methods are available for the simpler cases \citep[e.g., see][]{wright2024}.

We now consider the PGL defined in~\eqref{eq:7}. Firstly, we define
\begin{equation}\label{eq:11}
\bm\Sigma \equiv \left[
                     \begin{array}{cc}
                       \phi^2 & \rho\phi \\
                       \rho\phi & 1 \\
                     \end{array}
                   \right],
\end{equation}
where we adopted the constraint $\var(R\sin\Omega) = 1$ to ensure identifiability of the distribution of $\bm{S}$ \citep{wang2013,hernandez2017}. Secondly, we reparameterize $\bm{\delta} = (\delta_{1},\delta_{2})\tp$ where $\delta_{1} = \log(\phi)$ and $\delta_{2} = 1/2\log\{(1+\rho)/(1-\rho)\}$. The parameter to be estimated is $\bm{\eta} = (\bm{\theta}\tp, \bm{\delta}\tp, \zeta)\tp \in \mathbb{R}^{5}$. Then, for a sample $\bm{\Omega} = (\Omega_{1},\ldots,\Omega_{n}) \iid \mathcal{PGL}(\bm{\theta},\bm\Sigma,\alpha)$, the loglikelihood of $\bm{\eta}$ is given by
\begin{equation}\label{eq:12}
\ell(\bm{\eta}; \bm{\omega}) = \sum_{i=1}^{n} \log \int_{0}^{\infty} \left\{ \int_{0}^{\infty}  rf_{N}(r\bm{w}_{i}|v)\rds r\right\}g(v)\rds v,
\end{equation}
where $f_{N}(r\bm{w}|v)$ is the density of a bivariate normal in polar coordinates. Note that the inner integral has a closed-form solution, which is the density of a PN distribution \citep{hernandez2017}, that is
\begin{equation}\label{eq:13}
\int_{0}^{\infty}  rf_{N}(r\bm{w}_{i}|v)\rds r \equiv f_{PN}(\omega_{i}|v) = \frac{\exp\left(-\frac{1}{2v}\bm{\theta}\tp \bm{\Sigma}^{-1} \bm{\theta}\right)}{2 \pi (\bm{w}_{i}\tp \bm{\Sigma}^{-1} \bm{w}_{i})|\bm{\Sigma}|^{\frac{1}{2}}}\left[1+\frac{q_{i}\Phi(q_{i})}{\phi(q_{i})}\right],
\end{equation}
where $\bm{w}_{i} = (\cos\omega_{i},\sin\omega_{i})\tp$ and $q_{i}=\bm{w}_{i}\tp \bm{\Sigma}^{-1} \bm{\theta}(\bm{w}_{i}\tp \bm{\Sigma}^{-1}\bm{w}_{i}/v)^{-1/2}$, and $\phi$ and $\Phi$ are the standard normal density and cumulative distribution functions, respectively. As before, we propose using a GQ to solve the following (approximated) ML estimation problem
\begin{equation}\label{eq:14}
\hat{\bm{\eta}} = \argmax_{\bm{\eta}} \sum_{i=1}^{n} \log \left[\sum_{h=1}^{H} \left\{f_{PN}(\omega_{i}|v_{h}(\alpha))\right\}w_{h}(\alpha)\right].
\end{equation}
Like the optimization problem in~\eqref{eq:10}, the problem in~\eqref{eq:14} requires an iterative procedure. The estimated PGL density can be evaluated at any value of $\omega$ by plugging the solution $\hat{\bm{\eta}}$ in~\eqref{eq:7} and then using a numerical approximation to marginalize with respect to $r$ (e.g., with \texttt{stats::integrate} in \texttt{R}). (Of course, the values of the fitted density evaluated at the sample observations $\omega_{i}$, $i=1,\ldots,n$, are available as a by-product directly from the optimization procedure at convergence.) In contrast, the density in~\eqref{eq:6}, which depends on $\alpha$ only, is available analytically.

Note the symmetry between~\eqref{eq:10} and~\eqref{eq:14}, where $f(\cdot|v)$ is normal in the former equation and \emph{projected} normal in the latter. Integrating these two densities with respect to $v$ yields the GL and \emph{projected} GL, respectively.

\section{Numerical examples}\label{sec:3}
We ran a small simulation study in \texttt{R} \citep{R} to investigate the performance of the estimation method discussed in Section~\ref{sec:2}. In the first part of the study, we assessed the performance of GQ-based estimation for fitting the GL in~\eqref{eq:4} against two alternative methods, namely EM estimation and direct ML estimation, both implemented to fit GH distributions. To our knowledge, there are no studies that considered such a comparison. Since the goal was to fit the same distribution, but with different parameterizations \citep[e.g., refer to the package vignette in][for the GH parameterization]{ghyp}, we considered the mean squared error (MSE) of the mean and of the variance, rather than the MSE of the parameters per se. In the second part of the simulation study, we contrasted the newly proposed PGL with the PN and VM distributions. Unlike in the previous exercise, the goal here was to compare the goodness of fit of different models, hence we omitted the MSE. In both parts of the simulation, all optimization algorithms were allowed a generous maximum number of iterations (i.e., $5000$) to reduce the risk of premature termination, which could obscure the comparison.

\subsection{Simulation study part I}\label{sec:3.1}
In the first part of the simulation study, we compared GQ-based estimation with EM and direct ML estimation. The former is implemented in the \texttt{R} package \texttt{ghyp} \citep{ghyp} (specifically, the variance-gamma model \texttt{ghyp::fit.VGuv} and \texttt{ghype::fit.VGmv} for, respectively, univariate and multivariate variables), while the latter in the package \texttt{GeneralizedHyperbolic} \citep{GeneralizedHyperbolic}. These were used with their default optimization method (namely, Nelder-Mead in \textit{optim}). We considered four different flavors of our method: a Nelder-Mead optimization (\textit{optim}) with $H = 20$ quadrature points (GQ1); as GQ1, but with $H = 30$ (GQ2); a BFGS optimization (\textit{optim}) with $H = 20$ (GQ3); and a quasi-Newton optimization (\textit{nlminb}) with $H = 20$ (GQ4). To speed up computation, the GQ based on the gamma distribution as described in~\eqref{eq:10} and~\eqref{eq:14} was implemented in C++. We generated samples of three sizes $n \in \{30, 100, 500\}$ from three distributions: (i) $\mathcal{L}(0,1)$ as defined in~\eqref{eq:1}; (ii) $\mathcal{GL}(1,1,3,2)$ as defined in~\eqref{eq:3}; and (iii) $\mathcal{GL}_{d}(\bm{0}_{2},\bm\Sigma,(2,3)\tp,2)$ as defined in~\eqref{eq:4}, with
\begin{equation*}
\bm\Sigma = \left[
                     \begin{array}{cc}
                       2 & 1 \\
                       1 & 2 \\
                     \end{array}
                   \right].
\end{equation*}
Each of the 9 simulation scenarios (3 distributions and 3 sample sizes) was replicated $500$ times.

In Table~\ref{tab:A1} (Appendix), we report the MSE of the expected value and variance, the mean loglikelihood and elapsed time (seconds) to convergence, averaged across replications (excluding replications where the optimization failed to converge), and the proportion of convergence failures, while boxplots are shown in Figures~\ref{fig:A1}--\ref{fig:A3}. In summary, all methods perform similarly at the largest sample size ($n = 500$), except for direct ML that showed clear signs of struggle with the multivariate GL. At lower sample sizes ($n = 30$ or $n = 100$), GQ4 was in general very competitive in terms of MSE and convergence rates, especially when data were drawn from the GL (of particular note, GQ4 did not show erratic behavior as other methods did). In contrast, for univariate (classical) Laplace data, all methods performed similarly (although GQ1--GQ3 were more reliable in terms of convergence). Time-wise, GQ-based estimation was acceptable with data drawn from simpler distributions (Laplace and univariate GL), but otherwise less competitive than the other methods at $n = 500$. The only noteworthy difference when increasing the number of quadrature points (GQ1 vs GQ2) was a more reliable estimate of the scale of the GL.

\subsection{Simulation study part II}\label{sec:3.2}
In the second part of the simulation study, we generated samples of three sizes $n \in \{30, 100, 500\}$ from the distribution $\mathcal{PGL}((-2, 0),\bm{\Sigma},\alpha)$ as defined in~\eqref{eq:7}, with two settings: (i) $\bm{\Sigma} = \bm{I}_{2}$ and $\alpha = 10$ (this density is exemplified by the dashed line in Figure~\ref{fig:1}(a)); and (ii)
\begin{equation*}
\bm\Sigma = \left[
                     \begin{array}{cc}
                       30 & 4 \\
                       4 & 1 \\
                     \end{array}
                   \right]
\end{equation*}
and $\alpha = 0.5$ (this density is exemplified by the solid line in Figure~\ref{fig:1}(b)). We fitted three competing models: the PGL model, fitted by means of GQ1 and GQ4 as described in the previous section; the PN model, fitted by means of direct ML of \eqref{eq:13}; and the VM model, fitted by means of direct ML, which results in explicit formulas for the location and concentration parameters \citep{fisher1993}. Each of the 6 simulation scenarios (2 distributions and 3 sample sizes) was replicated $500$ times.

In Table~\ref{tab:A2} (Appendix), we report the mean loglikelihood and elapsed time (seconds) to convergence, averaged across replications (excluding replications where the optimization failed to converge), and the proportion of convergence failures, while boxplots are shown in Figures~~\ref{fig:A4}--\ref{fig:A5}. In summary, the three models exhibited similar goodness of fit in the first parameter setting of the GL. This was expected since the GL with large $\alpha$ yields a distribution that is compatible with a PN. Additionally, the unimodal shape resulting from this setting makes the VM an equally appropriate model to fit. In contrast, in the second parameter setting of the GL, neither the PN nor the VM was on par with the GL, with the GL giving a better fit than the other two. This result was also expected, as the PN cannot handle sharper peaks, while the VM cannot accommodate bimodality. However, while the inadequacy of the VM was evident even at the smallest sample size, the difference between the GL and the PN became more apparent at larger sizes. This is reasonable since accurate estimation of shape parameters requires larger samples than for other types of parameters. On the other hand, the behavior of the estimation methods GQ1 and, especially, GQ4 was somewhat unexpected. While the former never failed to converge in any of the settings, the latter struggled with a large proportion of convergence failures, though only in the first setting. We can only surmise that this was due to the flatter likelihood surface for which its derivative provided little information to feed to the quasi-Newton estimation, as confirmed by several \texttt{nlminb}'s \texttt{X-Convergence (3)} and \texttt{Relative Convergence (4)} messages (i.e, when the optimization process has stopped because the relative change in the parameters or the objective function is very small). Time-wise, GQ-based estimation was acceptable at lower sample sizes, but noticeably more demanding at $n = 500$, particularly for GQ1, as it is derivative-free and thus requires more function evaluations to converge. As part of a sensitivity analysis, we halved the number of quadrature points from $20$ to $10$ and re-ran the simulation with the same data generated in the first GL setting at $n = 500$. The mean time to convergence decreased from 9.80 to 5.00 seconds for GQ1, and from 6.38 to 4.59 seconds for GQ4, while the mean loglikelihood remained approximately the same.

\section{Final remarks}
We discussed the estimation of the GL distribution and of its projection onto the circle. The proposed \emph{projected} GL does not appear to have been considered in the literature of circular statistics.  Since this distribution lacks an analytical density, we considered a maximum likelihood estimation of its scale-mixture representation, coupled with numerical integration. We first explored this strategy for the (standard) GL distribution as the literature does not provide any guidance on whether GQ is competitive against more popular methods like EM and direct ML estimation. It should be stressed that the goal of this study was not to investigate computational issues and properties of estimation algorithms in general \citep[for such a study, we refer the reader to the recent work on the generalized hyperbolic distribution by][]{vanwyk2024}. The simulation revealed that GQ-based estimation performs consistently well. Indeed, it is reliable in general and has shown advantages over alternative estimation methods, especially with smaller samples. Still, optimizing the numerically integrated likelihood requires careful consideration, including testing different optimization algorithms and settings, as well as adjusting the number of quadrature points based on data size and characteristics. Finally, it should be borne in mind that fitting a projected GL is computationally more demanding than fitting a simpler distribution and may require a larger sample to distinguish its shape from others'.


\appendix

\renewcommand{\thefigure}{A\arabic{figure}}
\renewcommand{\thetable}{A\arabic{table}}
\setcounter{figure}{0}

\clearpage

\section*{Appendix}

\begin{table}
\caption{Mean squared error of the expected value (EV) and variance (var), mean loglikelihood and elapsed time (seconds) to convergence, averaged across replications, and proportion of convergence failures (X), for four different variants of the proposed estimation method based on Gaussian quadrature (GQ1-GQ4), expectation-maximization (EM) and direct optimization of the loglikelihood (direct ML), using data drawn from three distributions with three different sample sizes $n \in \{30, 100, 500\}$.}\label{tab:A1}
\centering
\begin{tabular}{@{}llrrrrrr@{}}
\toprule
 & $n$ & GQ1 & GQ2 & GQ3 & GQ4 & EM & Direct ML \\
\midrule
& & \multicolumn{6}{c}{\emph{Univariate classical Laplace}}\\
\cmidrule{3-8}
EV & 30 & 0.03 & 0.03 & 0.03 & 0.03 & 0.05 & 0.03 \\
  var &  & 0.19 & 0.21 & 0.19 & 0.19 & 1.41 & 0.14 \\
  loglik &  & $-$38.15 & $-$38.17 & $-$38.15 & $-$38.21 & $-$34.05 & $-$38.71 \\
  time &  & 0.07 & 0.10 & 0.07 & 0.03 & 0.05 & 0.00 \\
  X &  & 0.00 & 0.00 & 0.00 & 0.08 & 0.09 & 0.09 \\
  EV & 100 & 0.01 & 0.01 & 0.01 & 0.01 & 0.01 & 0.01 \\
  var &  & 0.05 & 0.05 & 0.05 & 0.05 & 0.06 & 0.04 \\
  loglik &  & $-$132.49 & $-$132.49 & $-$132.49 & $-$132.50 & $-$129.99 & $-$132.66 \\
  time &  & 0.08 & 0.11 & 0.08 & 0.05 & 0.04 & 0.01 \\
  X &  & 0.00 & 0.00 & 0.00 & 0.01 & 0.00 & 0.00 \\
  EV & 500 & 0.00 & 0.00 & 0.00 & 0.00 & 0.00 & 0.00 \\
  var &  & 0.01 & 0.01 & 0.01 & 0.01 & 0.01 & 0.01 \\
  loglik &  & $-$670.97 & $-$670.87 & $-$670.97 & $-$670.97 & $-$670.71 & $-$671.00 \\
  time &  & 0.27 & 0.41 & 0.27 & 0.28 & 0.07 & 0.02 \\
  X &  & 0.00 & 0.00 & 0.00 & 0.00 & 0.00 & 0.00 \\
\cmidrule{3-8}
& & \multicolumn{6}{c}{\emph{Univariate generalized Laplace}}\\
\cmidrule{3-8}
  EV & 30 & 0.96 & 0.97 & 0.96 & 1.00 & 2.56 & 1.02 \\
  var &  & 125.56 & 114.20 & 125.56 & 164.58 & 20357.61 & 189.23 \\
  loglik &  & $-$83.15 & $-$83.14 & $-$83.15 & $-$82.94 & $-$80.38 & $-$84.64 \\
  time &  & 0.15 & 0.22 & 0.15 & 0.06 & 0.08 & 0.01 \\
  X &  & 0.00 & 0.00 & 0.00 & 0.01 & 0.10 & 0.27 \\
  EV & 100 & 0.40 & 0.38 & 0.40 & 0.45 & 0.41 & 0.42 \\
  var &  & 51.01 & 46.43 & 51.01 & 70.61 & 62.15 & 123.33 \\
  loglik &  & $-$283.58 & $-$283.59 & $-$283.58 & $-$283.39 & $-$284.04 & $-$285.57 \\
  time &  & 0.21 & 0.32 & 0.21 & 0.08 & 0.04 & 0.01 \\
  X &  & 0.00 & 0.00 & 0.00 & 0.00 & 0.00 & 0.55 \\
  EV & 500 & 0.15 & 0.16 & 0.15 & 0.16 & 0.16 & 0.15 \\
  var &  & 24.51 & 24.67 & 24.51 & 26.88 & 24.84 & 76.73 \\
  loglik &  & $-$1425.48 & $-$1425.44 & $-$1425.48 & $-$1425.33 & $-$1425.26 & $-$1433.08 \\
  time &  & 0.68 & 1.01 & 0.69 & 0.26 & 0.07 & 0.02 \\
  X &  & 0.00 & 0.00 & 0.00 & 0.00 & 0.00 & 0.59 \\
\cmidrule{3-8}
& & \multicolumn{6}{c}{\emph{Multivariate generalized Laplace}}\\
\cmidrule{3-8}
  EV & 30 & 2.76 & 2.99 & 2.76 & 1.15 & 1.03 & 2.91 \\
  var &  & 4351.45 & 978.03 & 4351.45 & 197.72 & 1963.29 & 324.01 \\
  loglik &  & $-$137.91 & $-$137.92 & $-$137.91 & $-$134.08 & $-$124.50 & $-$158.46 \\
  time &  & 0.63 & 0.89 & 0.63 & 0.13 & 0.09 & 0.01 \\
  X &  & 0.03 & 0.02 & 0.03 & 0.00 & 0.31 & 0.40 \\
  EV & 100 & 0.59 & 0.86 & 0.59 & 0.39 & 0.35 & 2.32 \\
  var &  & 4166.27 & 2388.27 & 4166.27 & 60.03 & 276.00 & 172.83 \\
  loglik &  & $-$469.44 & $-$469.60 & $-$469.44 & $-$463.51 & $-$463.31 & $-$535.68 \\
  time &  & 1.14 & 1.64 & 1.14 & 0.23 & 0.22 & 0.01 \\
  X &  & 0.03 & 0.01 & 0.03 & 0.00 & 0.25 & 0.29 \\
  EV & 500 & 0.10 & 0.13 & 0.10 & 0.07 & 0.06 & 2.06 \\
  var &  & 40.37 & 40.71 & 40.37 & 9.55 & 8.83 & 105.11 \\
  loglik &  & $-$2354.90 & $-$2354.58 & $-$2354.90 & $-$2334.11 & $-$2334.09 & $-$2683.40 \\
  time &  & 6.84 & 10.02 & 6.80 & 1.59 & 0.77 & 0.04 \\
  X &  & 0.04 & 0.03 & 0.04 & 0.00 & 0.03 & 0.02 \\
\bottomrule
\end{tabular}
\end{table}

\clearpage

\begin{figure}[h]
    \centering
    \includegraphics[scale=0.5]{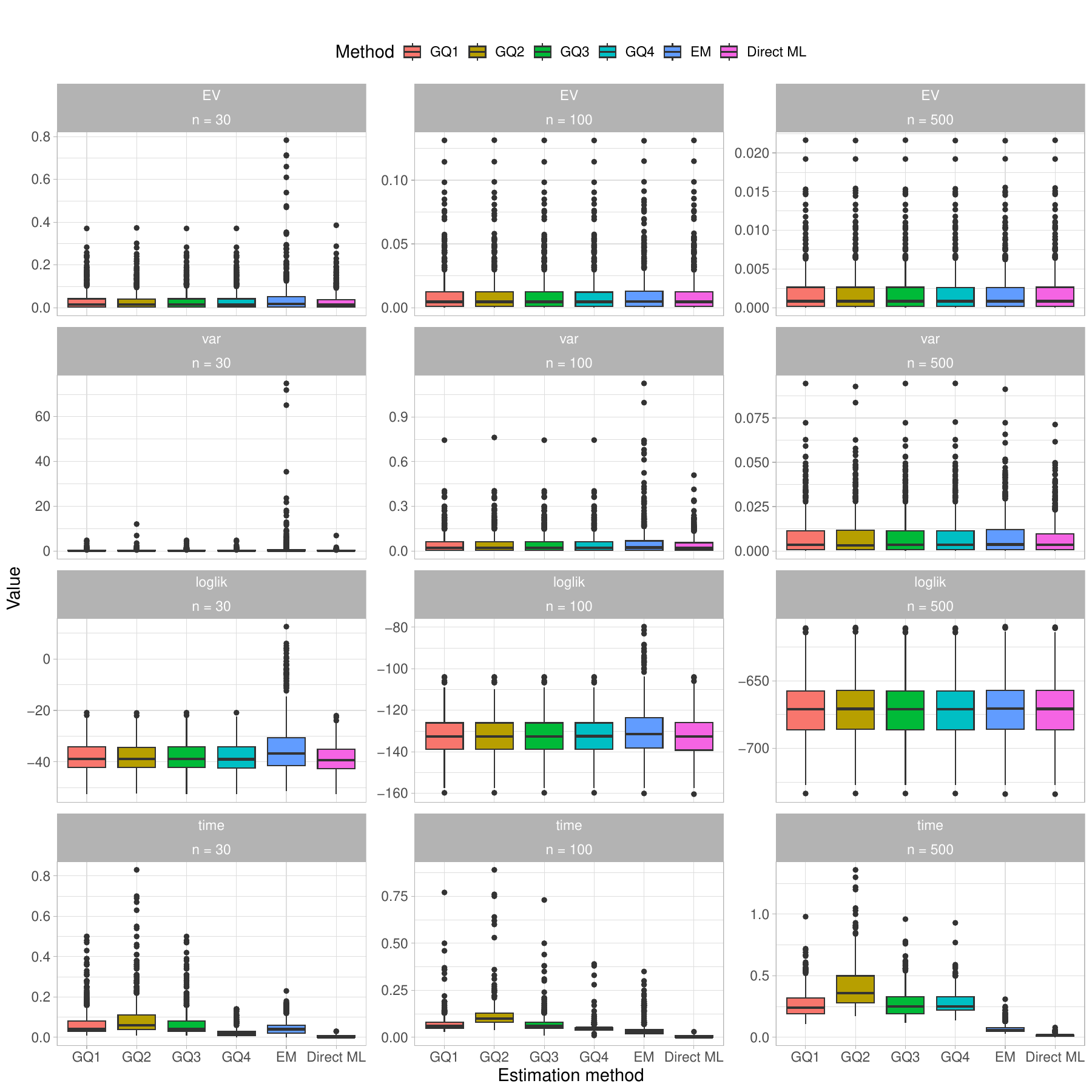}
    \caption{Boxplots of mean squared error of the expected value (EV) and variance (var), loglikelihood and elapsed time (seconds) to convergence, for four different variants of the proposed estimation method based on Gaussian quadrature (GQ1-GQ4), expectation-maximization (EM) and direct optimization of the loglikelihood (direct ML), using data drawn from the univariate Laplace with three different sample sizes $n \in \{30, 100, 500\}$.}
    \label{fig:A1}
\end{figure}

\clearpage

\begin{figure}[h]
    \centering
    \includegraphics[scale=0.5]{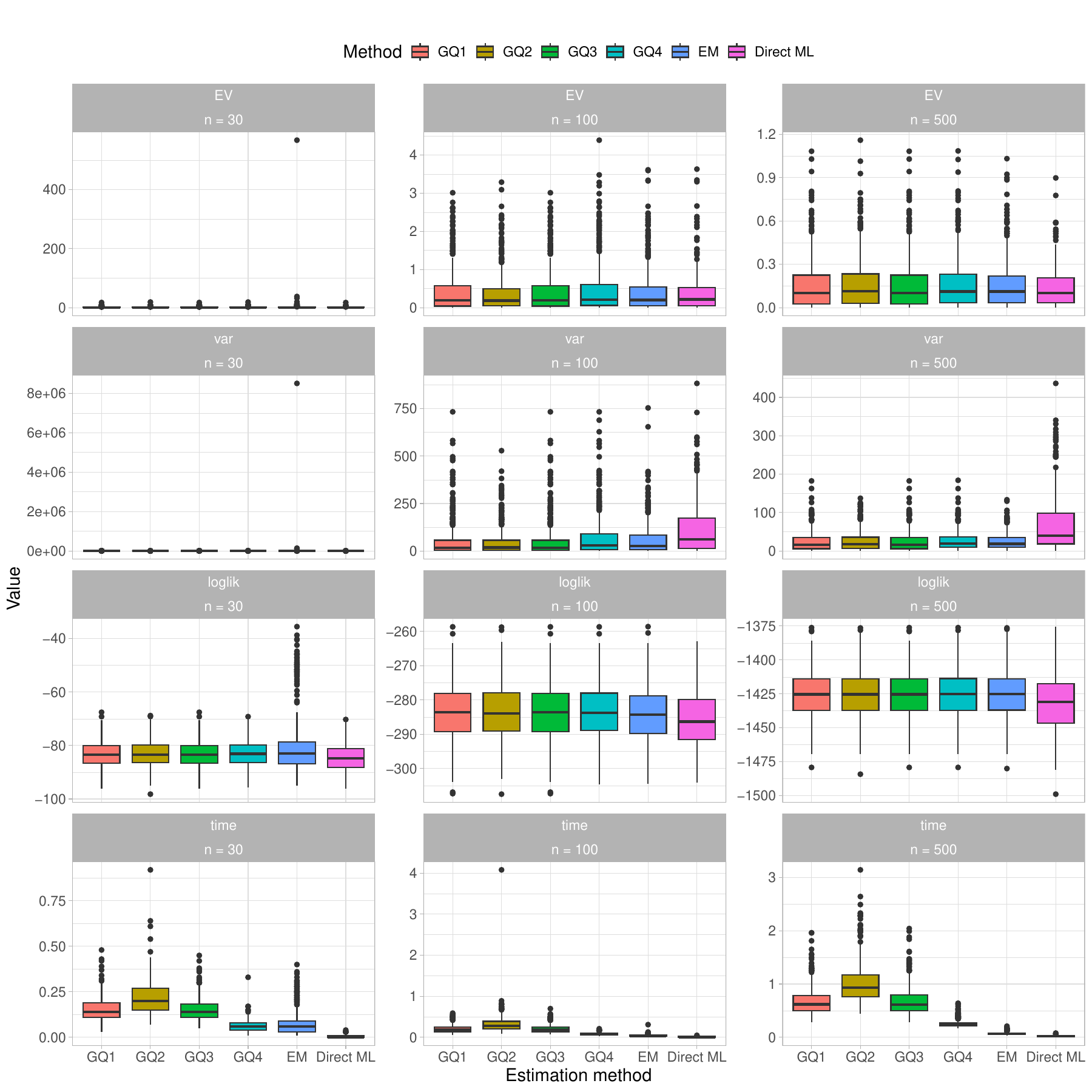}
    \caption{Boxplots of mean squared error of the expected value (EV) and variance (var), loglikelihood and elapsed time (seconds) to convergence, for four different variants of the proposed estimation method based on Gaussian quadrature (GQ1-GQ4), expectation-maximization (EM) and direct optimization of the loglikelihood (direct ML), using data drawn from the univariate generalized Laplace with three different sample sizes $n \in \{30, 100, 500\}$.}
    \label{fig:A2}
\end{figure}

\clearpage

\begin{figure}[h]
    \centering
    \includegraphics[scale=0.5]{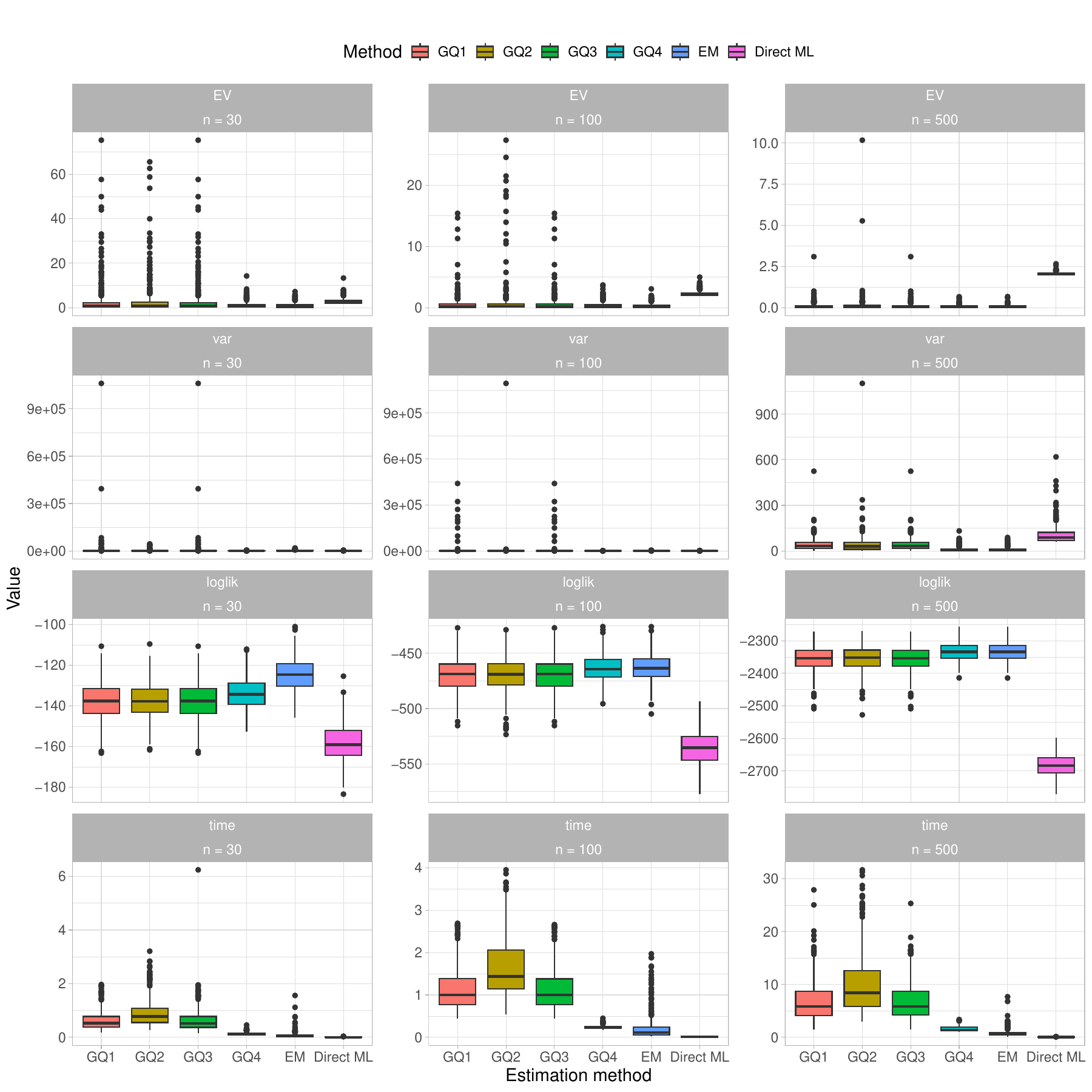}
    \caption{Boxplots of mean squared error of the expected value (EV) and variance (var), loglikelihood and elapsed time (seconds) to convergence, for four different variants of the proposed estimation method based on Gaussian quadrature (GQ1-GQ4), expectation-maximization (EM) and direct optimization of the loglikelihood (direct ML), using data drawn from the multivariate generalized Laplace with three different sample sizes $n \in \{30, 100, 500\}$.}
    \label{fig:A3}
\end{figure}

\clearpage

\begin{table}
\caption{Mean loglikelihood and elapsed time (seconds) to convergence, averaged across replications, and proportion of convergence failures (X), for the projected Laplace (PGL), projected normal (PN) and von Mises distributions, using data drawn from two PGL distributions with three different sample sizes $n \in \{30, 100, 500\}$.}\label{tab:A2}
\centering
\begin{tabular}{@{}llrrrr@{}}
\toprule
 & $n$ & PGL (GQ1) & PGL (GQ4) & PN & VM \\
\midrule
& & \multicolumn{4}{c}{\emph{Unimodal PGL with $\alpha = 10$ (approximately PN)}}\\
\cmidrule{3-6}
loglik & 30 & $-$48.26 & $-$48.50 & $-$48.32 & $-$49.38 \\
  time &  & 1.16 & 0.58 & 0.08 & 0.00 \\
  X &  & 0.00 & 0.38 & 0.00 & 0.00 \\
  loglik & 100 & $-$165.62 & $-$165.57 & $-$165.69 & $-$166.79 \\
  time &  & 2.49 & 1.45 & 0.16 & 0.00 \\
  X &  & 0.00 & 0.42 & 0.00 & 0.00 \\
  loglik & 500 & $-$837.40 & $-$837.08 & $-$837.50 & $-$838.81 \\
  time &  & 9.80 & 6.38 & 0.64 & 0.00 \\
  X &  & 0.00 & 0.45 & 0.00 & 0.00 \\
\cmidrule{3-6}
& & \multicolumn{4}{c}{\emph{Bimodal PGL with $\alpha = 0.5$}}\\
\cmidrule{3-6}
  loglik & 30 & $-$13.80 & $-$13.84 & $-$15.33 & $-$23.48 \\
  time &  & 0.81 & 0.24 & 0.07 & 0.00 \\
  X &  & 0.00 & 0.01 & 0.00 & 0.00 \\
  loglik & 100 & $-$50.36 & $-$50.29 & $-$55.99 & $-$81.47 \\
  time &  & 2.67 & 0.85 & 0.20 & 0.00 \\
  X &  & 0.00 & 0.00 & 0.00 & 0.00 \\
  loglik & 500 & $-$259.22 & $-$258.98 & $-$287.09 & $-$411.26 \\
  time &  & 12.92 & 4.70 & 0.92 & 0.00 \\
  X &  & 0.00 & 0.00 & 0.00 & 0.00 \\
\bottomrule
\end{tabular}
\end{table}

\clearpage

\begin{figure}[h!]
    \centering
    \includegraphics[scale=0.5]{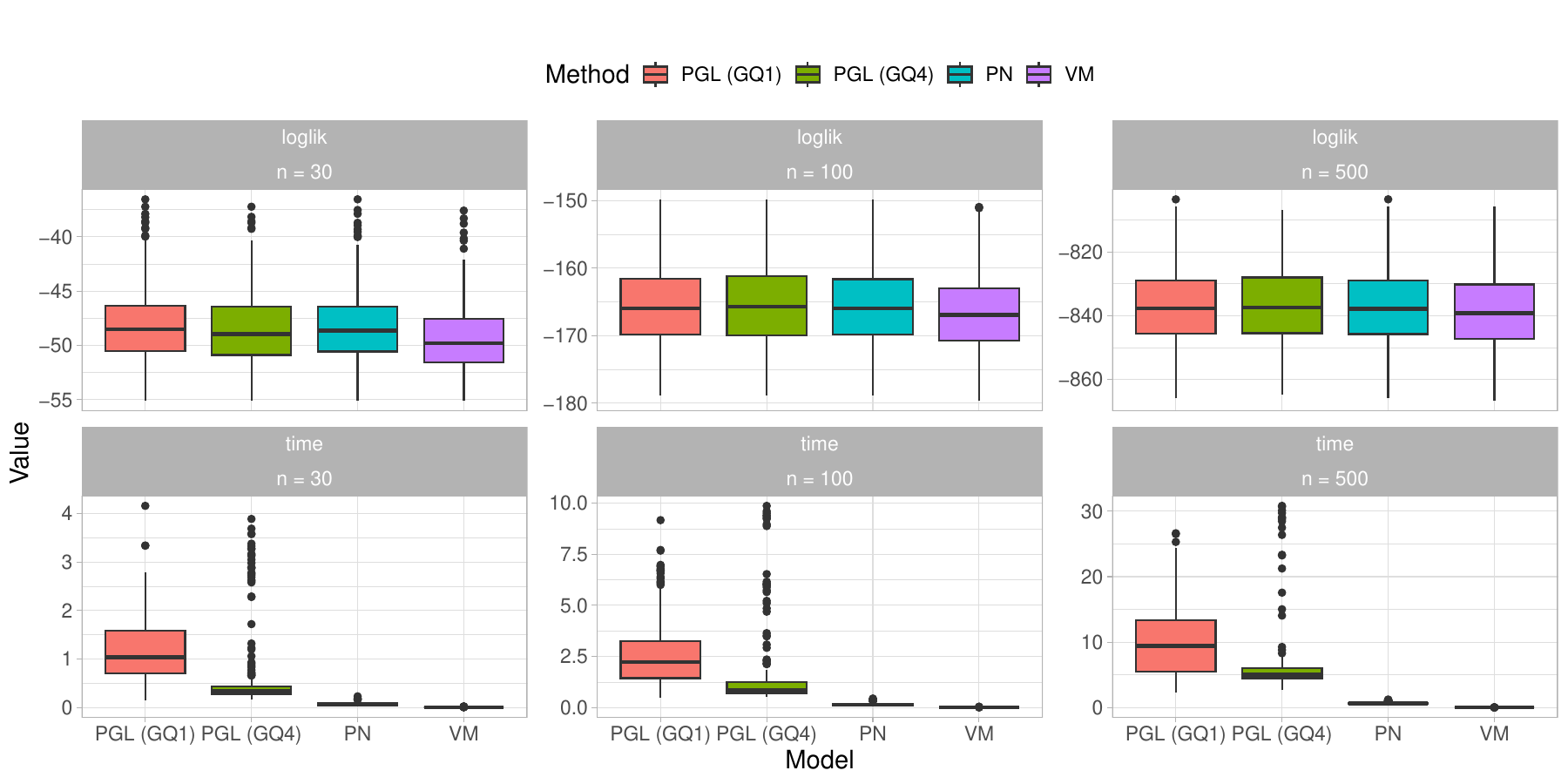}
    \caption{Boxplots of mean squared error of the expected value (EV) and variance (var), loglikelihood and elapsed time (seconds) to convergence, for the projected Laplace (PGL), projected normal (PN) and von Mises distributions, using data drawn from a unimodal PGL with $\alpha = 10$ (approximately PN) with three different sample sizes $n \in \{30, 100, 500\}$.}
    \label{fig:A4}
\end{figure}

\begin{figure}[h!]
    \centering
    \includegraphics[scale=0.5]{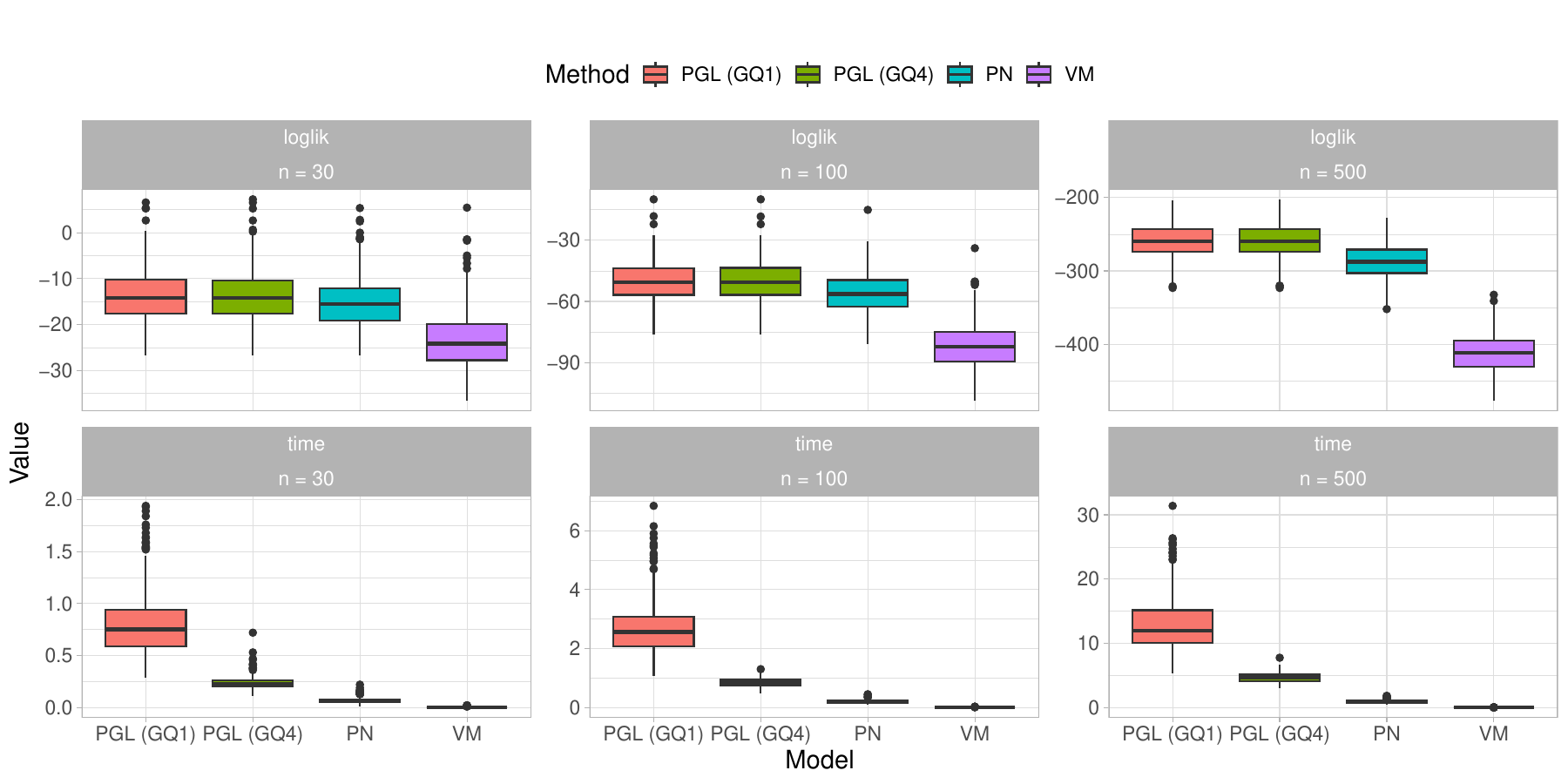}
    \caption{Boxplots of mean squared error of the expected value (EV) and variance (var), loglikelihood and elapsed time (seconds) to convergence, for the projected Laplace (PGL), projected normal (PN) and von Mises distributions, using data drawn from a bimodal PGL with $\alpha = 0.5$ with three different sample sizes $n \in \{30, 100, 500\}$.}
    \label{fig:A5}
\end{figure}

\clearpage


\end{document}